\documentclass[referee]{aa} 
\usepackage[latin1]{inputenc}
\usepackage[T1]{fontenc}
\usepackage{graphicx}
\usepackage{txfonts}
\usepackage{upgreek}
\usepackage{verbatim}
\usepackage{natbib}
\usepackage{color}

\newcommand{\lsol}{L$_\odot$\,}
\newcommand{\msol}{M$_\odot$\,}
\newcommand{\point}{\,}

\newcommand{\kms}{km$\point${s$^{-1}$}}

\newcommand{\msoly}{\msol$\point$yr$^{-1}$}

\newcommand{\cmcube}{cm$^{-3}$}

\newcommand{\ttp}[1]{$\times 10^{#1}$}

\newcommand{\Diff}[3]{\frac{d^{#3}{#1}}{d{#2}^{#3}}}

\newcommand{\new}[1]{#1}
\begin{document}
   \title{The Snow Border}

   \author{M.G. Marseille
          \inst{1}
          \and
          S. Cazaux
          \inst{2}
          }

   \institute{SRON Netherlands Institute for Space Research, Landleven 12,
9747AD
Groningen, The Netherlands\\
   email: \texttt{marseille@sron.nl}
         \and
	Kapteyn Astronomical Institute, PO Box 800, 9700 AV, Groningen, The
Netherlands\\
   email: \texttt{cazaux@astro.rug.nl}
	}

   \date{Received xxx; accepted xxx}

 
  \abstract
{
The study of the snow line is an important topic in several \new{domains} of astrophysics, and particularly for the evolution of proto-stellar environments and the formation of planets.
}
{
The formation of the first layer of ice on \new{carbon grains} requires low temperatures compared to the temperature of evaporation ($T>100$~K).  This asymmetry generates a zone in which bare and icy dust grains coexist. Our aim is to derive the proportion of bare grains around the theoretical snow line position for a typical low-mass protostellar disk and a massive protostar, and estimate the size of this mixing zone compared to the dust envelope size.
}
{
We use Monte-Carlo simulations to describe the formation time scales of ice mantles on bare grains in protostellar disks and massive protostars environments.
Then we analytically describe these two systems in terms of grain populations subject to infall and turbulence, and assume steady-state. 
Numerical applications use standard numbers obtained by previous observations or modelling of the astrophysical objects studied.
}
{
Our results show that there is an extended region beyond the snow line where icy and bare grains can coexist, in both proto-planetary disks and massive protostars.
This zone is not negligible compared to the total size of the objects: on the order of 0.4~AU for proto-planetary disks and 5400~AU for high-mass protostars.\rm{ Times to reach the steady-state are respectively estimated from $10^2$ to $10^5$~yr.}\rm 
}
{
The presence of a zone, a so-called \textit{snow border}, in which bare and icy grains coexist can have a major impact on our knowledge of protostellar environments. 
From a theoretical point of view, the progression of icy grains to bare grains as the temperature increases, could be a realistic way to model hot cores and hot corinos. 
Also, in this zone, the formation of planetesimals will require the coagulation of bare and icy grains.
Observationally, this zone allows high abundances of gas phase species at large scales, for massive protostars particularly, even at low temperatures (down to 50~K). This could be a critical point for the analysis of upcoming water observations by the Herschel Space Observatory. 
}


   \maketitle
%

\section{Introduction}

Inside the multiple tools that constrain the early phases of the formation of stars and planetary systems, water plays a major role. 
This molecule is one of the most abundant in the Universe, with $n_\mathrm{H_2O}/n_\mathrm{H_2} \sim 3 \times 10^{-4}$ in the hot parts of molecular clouds \citep{cernicharo1990,vandishoeck1996,harwitt1998,boonman2003} where it desorbs from the dust grain surface. 
Several observations of these hot parts have shown that water plays a major role in the formation of complex molecules \citep[\textit{e.g}][]{belloche2009}. 
This is due to the high dipole moment of this molecule that makes it active in both gas phase and surface chemistry. 
Water also acts as a host for complex molecules, keeping them in the icy mantle that covers dust grains. 
In addition, water plays a role in the energetic balance of protostellar environments \citep[\textit{e.g.}][]{ceccarelli1996,doty1997}.
The formation of planetesimals originates from the coagulation of icy dust grains: the formation of planets around stellar systems is triggered by the very first phases where dust grains stick to each other to form the bricks of rocky planets or the bulges of gaseous ones \citep{brauer2008}. 
Going a step further, water is the source of planetary oceans and is a necessary element for the appearance of life as known on the Earth.

Previous results on the sublimation of dust ices from solid surfaces were made in laboratories, showing that it occurs on short time-scales from 100~K up to 160~K \citep{sack1993,brown2007}, and that the chemical products enclosed in the ice mantle are released simultaneously. 
An astronomical example is given by complex molecules detected in hot corinos and hot molecular cores \citep[see][for a review]{cazaux2003,ceccarelli2007,beuther2007}. 
Whereas the sublimation of ices is roughly known, its condensation or direct formation on grains is poorly understood.  
Ice mantles originate from accretion of gas phase water onto dust grains \citep{tielens1982,cuppen2007}, or from direct formation of water via O and H chemistry on grain surfaces (\citealt{cazaux2010}). Experiments and theory show that on graphite surfaces, water molecules arrange themselves in clusters that have binding energies that increases with the size of the cluster \citep{lin2005,brown2007}. 
Once these clusters are created, water molecules are accreting onto themselves to form archipels from which the first layer of ice is made. 
This route is triggered by the condition that two molecules can meet each other on the surface to initiate the cluster. 
This process is limited if the temperature is too high, even for very high densities. 
In the case treated by \cite{cuppen2007}, $T\sim 10$~K and n$_\mathrm{H_2} = 10^4$~\cmcube, appearance of the first monolayer of ice takes at least $10^3$ years.
This might be significant compared to a disk or a massive protostar evolution time-scale (from $10^5$ to $10^7$~yr), and where density and temperature conditions are different.

At first, sublimation and condensation of water on dust particles has been considered to occur at the so-called snow line position, \textit{i.e.} where the temperature in the protostellar environment reaches temperatures around 100~K \citep[\textit{e.g.}][]{sasselov2000,lecar2006}. 
New results obtained on the formation of water ice show that the condensation of water is not optimal near this position, even impossible. 
Bare grains need to go back to a colder environment to create the very first ice layer and which allows the ice mantle to accrete very fast \citep{papoular2005}. 
Protostellar environments are known to be turbulent \citep[\textit{e.g.},][]{padoan2002,maclow2004}, leading to the idea of an extended zone where bare and icy dust grains can live together, creating the so-called \textit{snow border}. 
Its extension depends on the material transport time-scales compared to the condensation times. 
In this letter, we develop this idea for proto-stellar environments.
We first investigate the time scales required for the formation of the first layer of ice.
Then we derive standard equations that gives the mixing ratio $\zeta(r)$ between bare grains and the total amount of dust grains that can be expected around the classic snow line limit. 
We apply our model to a low-mass and a high-mass protostellar environment and conclude on the existence of a \textit{snow border}. 
At last, we discuss how the snow border could affect present and future observations or modelling of proto-stellar environments, and the formation of planetesimals.


\section{Water accretion modeling}

\rm {We use a step by step Monte Carlo simulation to describe the formation of water cluster on surfaces. The dust grains are divided with square lattices of $100 \times 100$ adsorption sites. Each site of the grid corresponds to a physisorbed site where water molecules can be bound with a determined binding energy. We consider that water molecules accrete from gas phase, and that the rate of accretion depends of the density and the velocity of the water molecules in the gas phase. This rate of accretion can be written as:
\begin{equation}
R_{acc} = n_{H_2O} \varv_{H_2O}  \sigma  S
\end{equation}
 where n$_{H_2O}$ and $\varv_{H_2O}$ are the densities and thermal velocities of the water species, $\sigma$ the cross section of the dust and $S$ is the sticking coefficient of the species with the dust. We consider $S=1$, meaning that when a water species arrive on a point of the grid, it becomes physisorbed. The water molecule that arrives on a location of the surface \new{is adsorbed with an energy $E_\mathrm{a}$ that depends on the numbers of water molecules in neighbouring sites (see Table~\ref{tab:energies}). Once the water is adsorbed on the surface, it can move randomly with a diffusion rate that can be written as 
 \begin{equation}
 R_\mathrm{diff} = \nu \exp\left(-\frac{E_\mathrm{diff}}{kT}\right)  
 \end{equation}
 where $E_\mathrm{diff} = 2/3 E_\mathrm{a}$ is the energy of the barrier between two sites \citep[from][]{cazaux2004}, $T$ the temperature of the surface and $\nu$ the oscillation factor of the water molecules (typically $10^{12}$~s$^{-1}$ for physisorbed species). Adsorbed molecules can also evaporates with a rate $R_\mathrm{evap}$ identical to $R_\mathrm{diff}$ except that $E_\mathrm{diff}$ is replaced $E_\mathrm{a}$. The molecules on the surface can execute a random walk on the surface that generates the formation of water clusters}. As two water molecules are in neighboring sites, a oxygen bound interaction link the two species, making the binding energy of the dimer higher than the binding energies of individual molecules. Then, other water molecules can join and increase the size of the cluster, and, as a result, increase the total binding energy (\citealt{gonzalez2007}). We used the binding energies calculated in \cite{lin2005}, as reported in Table~\ref{tab:energies}. A water cluster grows along the surface until it reaches six members. When the size of the cluster is larger than six molecules, the cluster grows in 3D \citep{gonzalez2007}, with a binding energy depending only on the number of water molecules which are close to the graphite surface \citep{lin2005}. Therefore, in our model, we limit the number of water that compose a cluster to 6 water molecules. The formation of cluster is therefore a competition between the evaporation of individual water molecules and the encounter of two molecules to initiate the cluster. \new{In this work, we only consider carbon surface. However, recent sutudies by \cite{goumans2009} show that the adsorption of water on silicate grains has very high binding energy (10\,000~K). Therefore, for these type of grains, we don't expect an assymetry between ice formation and evaporation.}\rm

\begin{table}
\caption{Adsorption energy (in kelvins) of water clusters as a function of the number of molecules present in the cluster, from \citealt{lin2005}.}
\label{tab:energies}
\begin{tabular}{lllllll}
&n=1&n=2&n=3&n=4&n=5&n=6 \\
\hline\hline
E$_a$ [K] & 1440 & 2400 & 3400 & 4600  & 5400 & 6400\\
\hline\hline
\end{tabular}
\end{table}
  

\section{The mixing ratio $\zeta(r)$}

In order to analytically describe the system, we consider a small zone around a radial position $r$ (see Fig~\ref{schema1}) with an extension $\Delta r$.
Inside this zone, all physical parameters are assumed to be constant. 
Going to infinitesimal variations in time and space then allow us to derive the equations describing the system evolution.
\rm{We follow the bare grain population, which evolves through the differences between the fluxes of grains moving away from the central star $K^+$ and the fluxes of grains going in the direction of the central star K$^-$. These fluxes are obtained from the description of the velocity fields inside the considered protostellar environments. }\rm
\rm{To study low-mass proto-planetary disks and massive protostar environments, we consider cylindrical and spherical descriptions of the system. For both of them,  we derive the mixing ratio $\zeta(r)$ between bare and total density of grains, as the system reaches steady state. Thus this mixing ratio describes the snow border itself. The first subsection gives the analytic description of the systems and the velocity fields acting on them, then the next subsection discuss the time scales and feasibility of reaching steady state.}\rm 


\begin{figure}[t]

 \begin{center}
 \includegraphics[width=200pt]{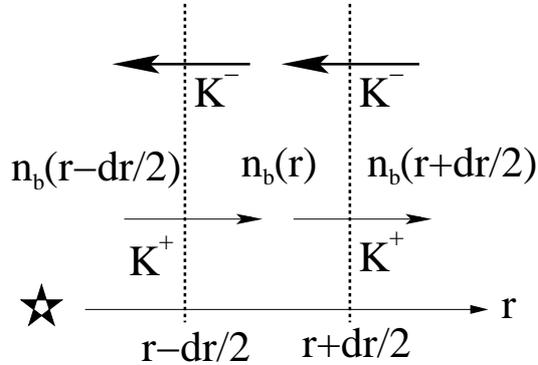}
\end{center}
\caption{System description used for the analytic description of the dust grain movements. We consider very small elements of width $\Delta r$ in which density of bare grains are constant with the value $n_b(r)$. Between a small time range $\Delta t$, due to turbulences, diffusion of bare grains ($n_b(r)$) is possible, triggered by $K^\pm(r)$ fluxes (see Sect.~3).}
\label{schema1}
\end{figure}

\begin{figure}[t]
\begin{center}
 \includegraphics[width=240pt]{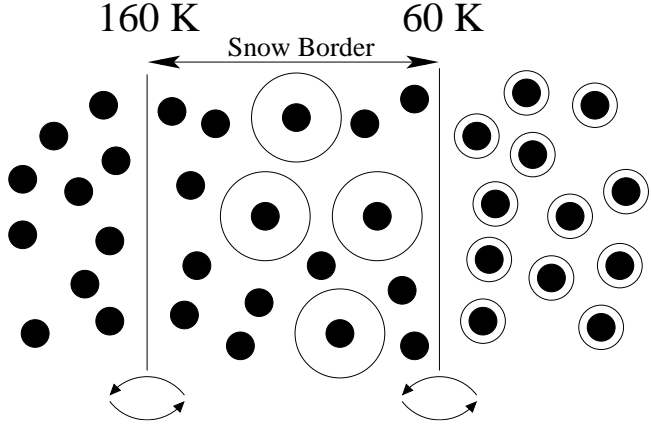}
\end{center}
 \caption{\rm{Overview of the snow border in a proto-stellar environment. At a certain distance from the central star, there is a zone where bare and icy grains coexist. This zone is located between two regions: maximum temperature region at which ices evaporate (here 160K) and a minimum temperature region at which the first layer of ice can form (here 60K). Turbulences allow grain and gas to mix}\rm.}
 \label{fig:overview}
\end{figure}

\subsection{Velocity fields}

We consider three main components acting on the velocity field: the thermal velocity $\varv_\mathrm{\theta}$, the turbulent velocity $\varv_\mathrm{t}$ and the accretion velocity $\varv_\mathrm{acc}$. 
The two first components are assumed to be isotropic and can be averaged geometrically to form a composite velocity $\varv_\mathrm{T}$:
\begin{equation}
\varv_\mathrm{T} = \left[\frac{2k_\mathrm{B}T}{m} +
\varv^2_\mathrm{t}\right]^{1/2}
\end{equation}  
where $T$ is the dust temperature, $k_\mathrm{B}$ is the Boltzmann constant, $m$ is the average mass of a dust grain and $\varv_\mathrm{t}$ can be determined with molecular line observations. 
The corresponding velocity field can thus be assumed to follow a Maxwell-Boltzmann distribution of the form:
\begin{equation}
f(\varv) = \frac{4 a^3}{\pi^{1/2}} \varv^2 e^{-a^2 \varv^2}
\end{equation}
where $a=1/\varv_\mathrm{T}$. 
In our case, we are interested in the radial distribution of the velocities which allow the transport of grains to higher or lower temperatures. 
In proto-stellar environments, where the accretion of materials onto the central protostar is done radially through disks or shells, the radial field $f_r(\varv)$ follows the form:
\begin{equation}
f_r(\varv_r) = \frac{a}{\pi^{1/2}} e^{-a^2 \varv_r^2}
\end{equation}
This velocity distribution is shifted by the velocity field induced by the accretion. 
Knowing the accretion rate $\dot{M}$ of the protostellar object permits to deduce $\varv_\mathrm{acc}$:
\begin{eqnarray}
\varv_\mathrm{acc}(r) & = & -\frac{\dot{M}}{4\pi r \mu m_p n(r) H(r)}, \quad \mbox{ cylindrical, } \\
 & = & -\frac{\dot{M}}{4\pi r^2 \mu m_p n(r)}, \quad \mbox{ spherical, }
\end{eqnarray}
where $\mu$ is the average mass of a gas particle, $m_p$ the proton mass, $n(r)$ the H$_2$ density and $H(r)$ the height scale of the disk at the distance $r$.
We note that $\varv_\mathrm{T}$ also depends on $r$ since it contains a thermal component varying with $T(r)$. 
The new radial velocity distribution $f_r^*(\varv_r)$ is then obtained by shifting the initial distribution:
\begin{eqnarray}
f^*_r(\varv_r) & = & f_r(\varv_r-\varv_\mathrm{acc})
\end{eqnarray}
This result allows us to define an average velocity for dust grains travelling to lower radius, denoted $\left<\varv^-\right>$, and another one for the opposite trend, called $\left<\varv^+\right>$:
\begin{eqnarray}
\left<\varv^-\right>(r) & = & \displaystyle\int_{-\infty}^{0} \varv f^*_r(\varv_r) d\varv \\
\left<\varv^+\right>(r) & = & \displaystyle\int_{0}^{+\infty} \varv f^*_r(\varv_r) d\varv
\end{eqnarray}
giving
\begin{eqnarray}
\left<\varv^-\right>(r) & = & \frac{\varv_\mathrm{acc}}{2}\left[1-\mathrm{Erf}\left(a\varv_\mathrm{acc}\right)\right]-\frac{e^{-a^2\varv^2_\mathrm{acc}}}{2a\pi^{1/2}} \\
\left<\varv^+\right>(r) & = & \frac{\varv_\mathrm{acc}}{2}\left[1+\mathrm{Erf}\left(a\varv_\mathrm{acc}\right)\right]+\frac{e^{-a^2\varv^2_\mathrm{acc}}}{2a\pi^{1/2}}
\end{eqnarray}
\rm{where $\mathrm{Erf(x)}$ is the standard error function.
Now that the mean velocities of inward or outward motions are known, we derive the fractions of dust grains $N^\pm(r)$ having positive or negative velocities. They are given by:
\begin{eqnarray}
N^-(r) & = & \displaystyle\int_{-\infty}^{0} f^*_r(\varv_r) d\varv  \\
N^+(r) & = & \displaystyle\int_{0}^{+\infty} f^*_r(\varv_r) d\varv
\end{eqnarray}
hence
\begin{eqnarray}
N^-(r) & = & \frac{1}{2}\left[1 - \mathrm{Erf}(a(r)\varv_\mathrm{acc}(r))\right] \\
N^+(r) & = & \frac{1}{2}\left[1 + \mathrm{Erf}(a(r)\varv_\mathrm{acc}(r))\right].
\end{eqnarray}
These numbers allow to derive flux densities of dust grains $K^\pm$ at a certain position $r$ in the model that we define. These fux densities depend on the geometry considered:
\begin{eqnarray}
K^\pm(r) & = & \pm 2 \pi r H(r) \left<\varv^\pm\right> N^\pm(r), \quad \mbox{ cylindrical, } \\
         & = & \pm \pi r^2 \left<\varv^\pm\right> N^\pm(r), \quad \mbox{ spherical, }
\end{eqnarray}
where the sign $\pm$ in front of the expressions are introduced to make the value of flux densities always positive. Indeed, from its definition, $\left<\varv^-\right>(r)$ has negative values.}\rm

\subsection{Radial dependence: $\zeta(r)$}

We consider the variation in bare grain density $n_b(r)$ inside a shell of width $\Delta r$ around the position $r$ (see Fig.~\ref{fig:schema}). Bare grains  move from $r$ with flux densities $K^\pm (r)$ or disappear into icy form with a speed $k_c(r)$ (in grains.s$^{-1}$). \rm{Determining this last parameter in then possible through the modeling we described in previous section, and empirical formul\ae{} will be given for each case studied in the next section. Indeed, $k_c$ depends on $r$ as it is linked to the local conditions, \textit{i.e.} $n(r)$ and $T(r)$, and can present huge variations depending on the proto-stellar environment studied.
 
Variations in the total number of bare grains $N_b(r)$  between $r-\Delta r/2$ and $r+\Delta r/2$ are then described by:
\begin{eqnarray}
\label{eq:ngrains}
N_b(r,t+\Delta t) & = & N_b(r,t) -  N_\mathrm{frozen} + N_\mathrm{in} - N_\mathrm{out}
\end{eqnarray}
where $N_\mathrm{frozen}$ is the amount of bare dust grains becoming frozen, \textit{i.e.} having formed 10~\% of the first ice layer, per time unit. This number is related to $k_c$ by:
\begin{eqnarray}
N_\mathrm{frozen} & = & k_c(r) N_b(r,t) \Delta t
\end{eqnarray}
but can also be related to the local density introducing the volume element depending on the model geometry:
\begin{eqnarray}
N_\mathrm{frozen} & = & K_c(r) n_b(r,t) \Delta r \Delta t
\end{eqnarray}
hence
\begin{eqnarray}
K_c(r) & = & 4\pi r H(r) k_c(r), \quad \mbox{ cylindrical, } \\
       & = & 2\pi r^2 k_c(r), \quad \mbox{ spherical.}
\end{eqnarray}
Other terms in Eq.~\ref{eq:ngrains}, $N_\mathrm{in}$ and $N_\mathrm{out}$, refer to the incoming and outcoming numbers of grains due to velocity fields described previously. These fields are inducing fluxes that depend on the geometry considered:
\begin{eqnarray}
K^\pm(r) & = & \pm 2 \pi r H(r) \left<\varv^\pm\right> N^\pm(r), \quad \mbox{cylindrical,} \\
K^\pm(r) & = & \pm \pi r^2 \left<\varv^\pm\right> N^\pm(r), \quad \mbox{spherical,}
\end{eqnarray}
giving the following expressions:
\begin{eqnarray}
N_\mathrm{in} & = & K^+\left(r-\frac{\Delta r}{2}\right)n_b\left(r-\frac{\Delta r}{2}\right)\Delta t \nonumber \\ 
              & + & K^-\left(r+\frac{\Delta r}{2}\right)n_b\left(r+\frac{\Delta r}{2}\right)\Delta t \\
N_\mathrm{out} & = & K^+(r)n_b(r)\Delta t + K^-(r)n_b(r)\Delta t.
\end{eqnarray}
Going to the limits when $\Delta t \rightarrow 0 $ and $ \Delta r \rightarrow 0$ yields the differential equation
\begin{eqnarray}
\Diff{}{r}{}\left(Jn_b\right) - n_b K_c = 0
\end{eqnarray}
with
\begin{eqnarray}
J(r) & = & K^+(r) - K^-(r)
\end{eqnarray}
which can be solved numerically with the boundary condition that all the ice mantles are totally sublimated beyond 160~K ($n_b(r_s)=n(r_s)$). 
This temperature corresponds to the desorption of the most strongly bound ice structures that can be created on dust (\textit{i.e.} thick water ice, \citealt{collings2004}, on graphite surface, \citealt{brown2007}). The snow border extend can then be derived through the ratio $\zeta(r)$ given by:
\begin{eqnarray}
\zeta(r) & = & \frac{n_b(r)}{n(r)}.
\end{eqnarray}
We then assume in our study that the snow border extension $\Delta R$ corresponds to the distance between the ice mantles sublimation radius and the radius where $\zeta(r) = 0.5$.}\rm 

\subsection{Time estimate to reach steady-state : $\tau$}

\rm{The snow border extend refers to the steady-state of the system. In this sense, one must consider the time to reach this equilibrium. In our study, an estimate of this time is critical, as proto-stellar environments have a limited lifetime linked to the accretion rate and the final star mass which is formed. As the system must be stable on a given period to fit our assumptions on the density and temperature distributions, the snow border idea makes sense only if its growth is fast enough.

To roughly estimate the time $\tau$ needed to reach the steady-state, we focus on the mean velocity of grains going to the external part of the proto-stellar environments, \textit{i.e.} $\left<\varv^+\right>$. An estimate of $\tau$ is then given by:
\begin{eqnarray}
\tau \sim \displaystyle\int_{r_s}^{r_s+\Delta R} \frac{dr}{\left<\varv^+\right>(r)}.
\end{eqnarray}
As this value depends critically on the diffusion length scale in real systems, this estimate assumes that turbulences are intermediate scale motions compared to the size of the proto-stellar environment. This makes this first estimate of $\tau$ a lower limit to reach the steady-state, as the diffusion speed can critically increase if the turbulence typical length scale is very small. In our case, disk-like environments are the most difficult to treat, and $\tau$ estimates we obtain can be realistic restricting the model to the mid-plane, where turbulences are near to the thermal width, and gas is coupled with grains. For high-mass protostars, both study of molecular clouds dynamics and observation gives a correlation length between $0.05$~pc and $0.1$~pc, \textit{i.e.} the typical size of a massive dense core \citep{miesch1994,goodman1998,ossenkopf2001}. We can then be relatively confident with this first estimation, which is also coherent with the first-guess analytic approach we have in this study. 

\section{Astrophysical applications}

We first apply the model described previously to two particular proto-stellar environment where the snow border may have an important extend, hence impact. We choose a typical proto-planetary disk and a massive dense core, in which low-mass and high-mass star formation occur. This goal of this first step is to disantangle the fundamental parameters that trigger the snow border extend and growth, before going to a general test on the model covering a grid of these parameters.}\rm 

\subsection{Proto-planetary Disks}

The snow line position is very important to constrain the formation of planetesimals in low-mass star forming regions, \rm{as current models treating this topic take into account bare-bare or icy-icy grain collisions only. During the formation of low-mass stars, accretion of materials on the central protostar is done through a rotating gaseous and dusty disk, which is the precursor of the planetary system. 
As a first test case for our modeling, we use for our application a model of a proto-planetary disk around a T-Tauri star by \cite{woitke2009}. Turbulence, observed with molecular line emissions \rm{towards a disk surrounding the Herbig Ae star HD 34282, is of the order of $0.1$~\kms\ \citep[\textit{e.g.}][]{pietu2003}. This value, has been estimated for a Herbig AE disk that is more massive and somewhat hotter than the TTauri disk considered in this study. Since such turbulent bandwidths have not been detected towards TTtauri disk, we consider $0.1$~\kms\ as an upper limit}\rm. We also consider a mass accretion rate in this object of $10^{-6}$~\msoly, \rm{which is also an upper limit  (\citealt{hueso2005})}\rm. According to the turbulence limitation we discussed in the previous section, we restrict the snow border extend and growth in the mid-plane of the accretion disc. This also implies a very low UV field for the modeling of the water accretion on carboneous dust grains.

In the mid-plane, H$_2$ density is known to follow a power-law of the form:
\begin{eqnarray}
n(r) & = & n_0 \left(\frac{r}{r_0}\right)^p
\label{eq:power-law}
\end{eqnarray}
where $n_0$ is the density at the reference radius $r_0$ and $p$ a power-law index, equal to $-2.45$ in the model by \cite{woitke2009}. The disk height scale $H(r)$ is proportionnal to $r^2$ and the temperature $T(r)$ is proportionnal to $r^{-0.25}$, both of them following again a power-law similar to the one indicated for the density. Values of reference at $r_0 = 1$~AU are $n_0 = 10^{14}$~\cmcube, $T_0 = 130$~K and $H_0 = 0.1$~AU. Conversion to dust grain density assumes a standard MRN distribution (\citealt{mathis1977}) with gas-to-dust ratio of 100. 

Five density-temperature couples corresponding to the proto-planetary-disk are used to derive an empirical formula giving the condensation speed $k_c$, from the results obtained by the grain surface modeling. The water abundance assumed, relative to H$_2$, is $10^{-4}$ according to atomic universal abundances. Ice mantle growth rates obtained are given in Fig.~\ref{fig:kc}, and the corresponding condensation speeds are given in details in Table~\ref{tab:wac-pp}. Resulting formula for the proto-stellar disk case in the mid-plane is:
\begin{eqnarray}
k_c(r) & \simeq & 6.65 \times 10^{-3} e^{-0.12\times T(r)} \mathrm{[s^{-1}]}.
\end{eqnarray}

\begin{table}[t]
  \begin{center}
    \begin{tabular}{ccc|ccc}
    \multicolumn{3}{c}{Proto-planetary disk} & \multicolumn{3}{c}{HMPO} \\
$T$ [K]	& $n_\mathrm{H_2O}$ [\cmcube] 	& $k_c$ [s$^{-1}$] & $T$ [K]	& $n_\mathrm{H_2O}$ [\cmcube] 	& $k_c$ [s$^{-1}$] \\
\hline
\hline
60	&	5.1\ttp{6}	& 3.3\ttp{-6} & 20	& 8.6\ttp{-7} & 5.9\ttp{-10}  \\
70	&	2.3\ttp{7}	& 2.4\ttp{-6} & 30	& 2.0\ttp{-6} & 1.7\ttp{-9} \\
80	&	8.6\ttp{7}	& 5.9\ttp{-7} & 40	& 3.5\ttp{-6} & 1.4\ttp{-11} \\
90	&	2.7\ttp{8}	& 2.1\ttp{-7} & 50	& 5.5\ttp{-6} & 4.2\ttp{-13} \\
100	&	7.6\ttp{8}	& 3.0\ttp{-8} & & & \\
\hline
    \end{tabular}
  \end{center}
\caption{Results of the water accretion modeling on carbenous grains in the mid-plane of a proto-planetary disk (left) and in high-mass proto-stellar object (right).}
\label{tab:wac-pp} 
\end{table}

\begin{figure}[t]
\begin{center}
 \includegraphics[width=240pt,angle=-90]{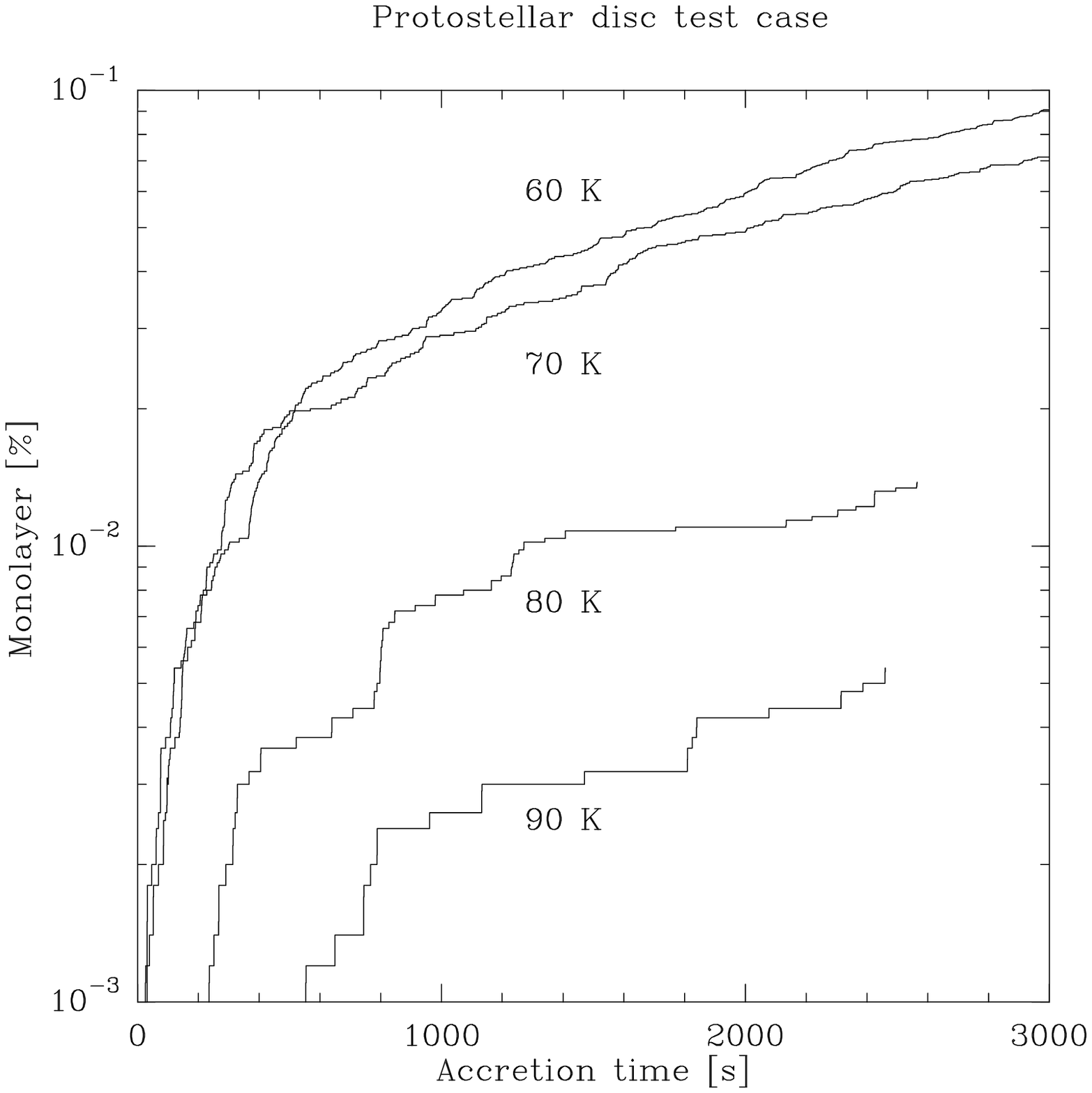}
 \includegraphics[width=220pt,angle=-90]{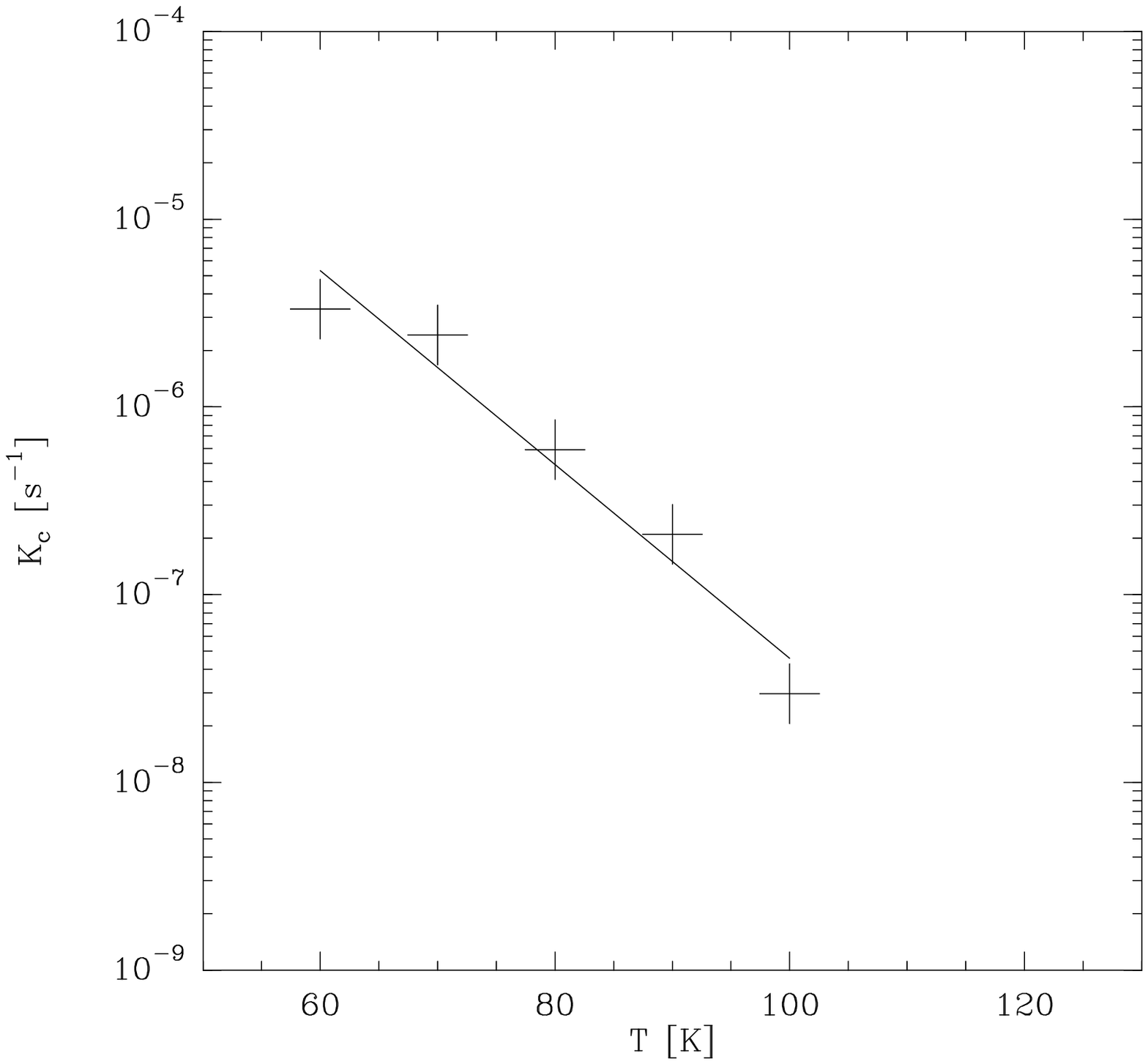}
\end{center}
 \caption{Plot of the bare-to-icy grain transformation rate $k_c$ as function of the temperature in the proto-planetary disk considered.}
 \label{fig:kc}
\end{figure}

In this case, the limit where $T>160$~K is $r_\mathrm{evap} = 0.43$~AU. 
The \textit{snow border} extension is shown in Fig.~\ref{fig:ppdisk}. 
The result shows an extended region where the mixing ratio is significant. 
The domain where $\zeta > 0.5$ is spread over $\Delta{r} \sim 0.2$~AU, that corresponds to $T \sim 135$~K. 
A first conclusion is that the \textit{snow border} does not reach its maximally possible extent, because diffusion of bare grains is opposed by the accretion velocity field and re-formation of ice mantles. Evaluation of the diffusion time $\tau \sim \Delta{r}/\left<v^+\right>(r_\mathrm{evap})$ gives the steady-state value $\tau \sim 10^2$~yr, \textit{i.e.} shorter than the disk lifetime.

\begin{figure}[t]
 \begin{center}
 \includegraphics[width=240pt,angle=-90]{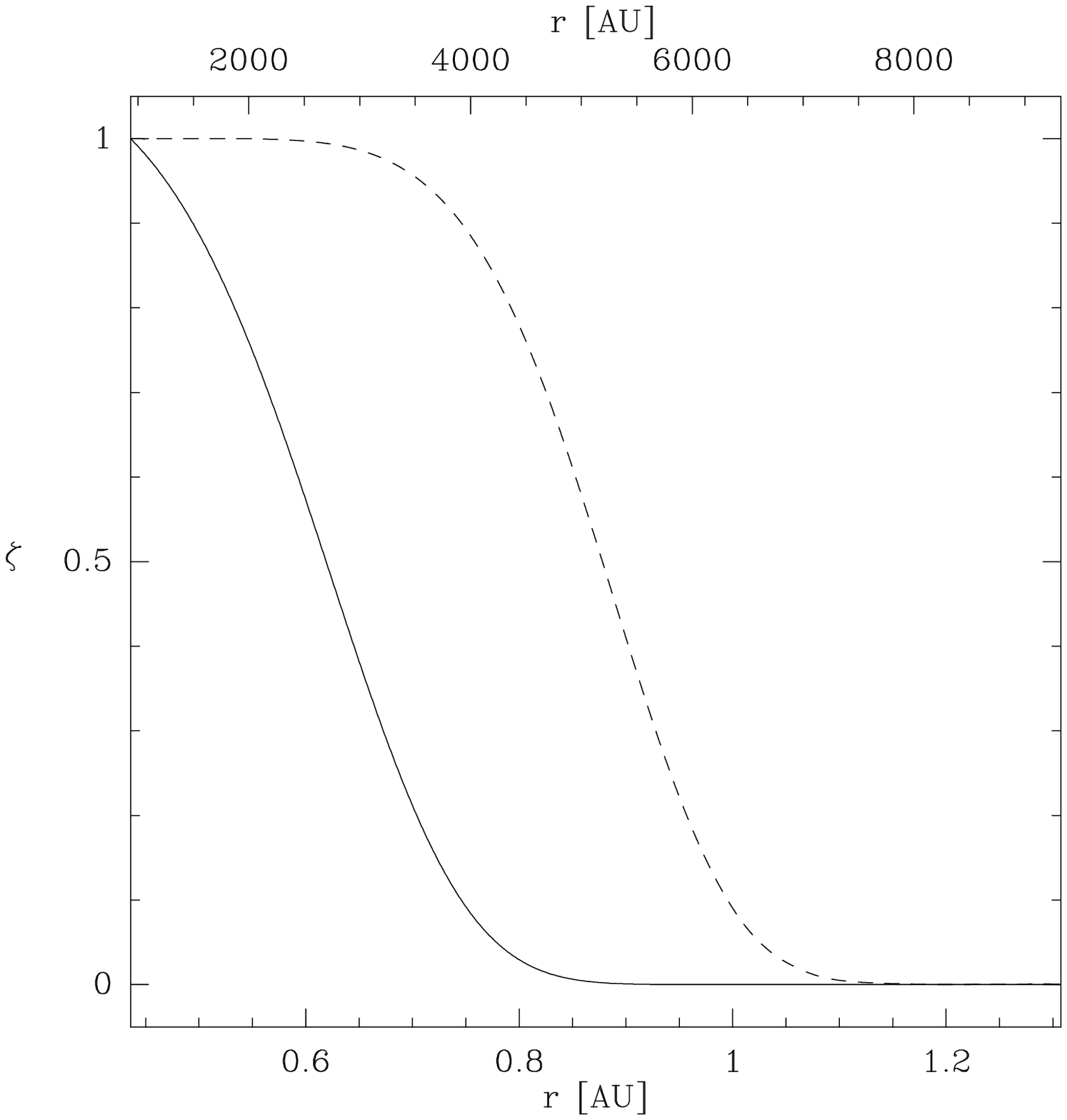}
 \end{center}
\caption{Plot of the mixing ratio $\zeta(r)$ as a function of the radius for: T-Tauri proto-planetary disk models (solid line, bottom labels), massive protostars (dashed line, upper label).}
\label{fig:ppdisk} 
\end{figure}

\subsection{High-Mass Protostars}

The \textit{snow border} is particularly interesting to follow in the case of massive protostars, where strong turbulence is observed ($\varv_\mathrm{t} > 1$~\kms) in addition to high accretion rates ($\dot{M} > 10^{-5}$~\msoly). 
Furthermore, due to their high luminosities ($L>10^4$~\lsol), the initial snow line position in such objects is shifted to higher distances. At this location, gas density drops by orders of magnitude compared to low-mass protostars ($10^5$ to $10^6$ times lower) which should have a strong effect on the formation of the first layer of ice. 

In this study we consider a turbulent velocity field value of $\sim 1$~\kms, as shown by molecular line width measurements and modeling. We also include a mass accretion rate of 1\ttp{-3}~\msoly\ as required for the formation of high-mass stars, and consider the spherically symmetric expressions for $K^\pm$ and $k_c$. \rm{The early stages of high-mass protostars have been modeled considering a 1D shape ( \citealt{marseille2008}), which allow us to derive the gas density and temperature as function of the distance from the central star}\rm. The gas density distribution used follows a power-law similar to Eq.~\ref{eq:power-law}, with $r_0 = 21$~AU, $n_0 = 2.8$\ttp{9}~\cmcube\ and $p=-1.2$, giving a temperature distribution following the same behavior, with $T_0 = 1500$~K and $p=-0.59$, according to the modeling of the IRAS~18151$-$1208(MM1) massive dense core in \cite{marseille2008}. 

\rm{As for the case of proto-planetary disk, we define several density-temperature couples that corresponds to high mass protostellar objects. For each of these couples (see Table~\ref{tab:wac-pp}), we calculate the time to form the first layer of ice using our grain surface modeling. With these calculations, we can derive an empirical formula giving the condensation speed $k_c$. In a similar way than for proto-planetary disks, we consider the water abundance of  $10^{-4}$ relative to H$_2$. Ice mantle growth rates obtained are given in Fig.~\ref{fig:kc-mass}, and the corresponding condensation speeds are given in details in Table~\ref{tab:wac-pp}. Resulting formula for high mass protostellar objects is:}\rm
\begin{eqnarray}
k_c(r) & \simeq & 2.97 \times 10^{-8} e^{-0.23\times T(r)} \mathrm{\,[s^{-1}]}.
\end{eqnarray}

The derived mixing ratio $\zeta(r)$ shows that the domain where $\zeta > 0.5$ is spread over $\sim 5000$~AU (see Fig.~\ref{fig:ppdisk}), reaching $T \sim 55$~K. However, the diffusion time scale is much longer in this case ($\tau \sim 10^5$~yr), indicating that the system may not have time to reach this steady state completely. In the case of high-mass protostars is then very extended because of the high turbulences and the low densities in the region where water freezes-out onto dust.

\begin{figure}[t]
\begin{center}
 \includegraphics[width=240pt,angle=-90]{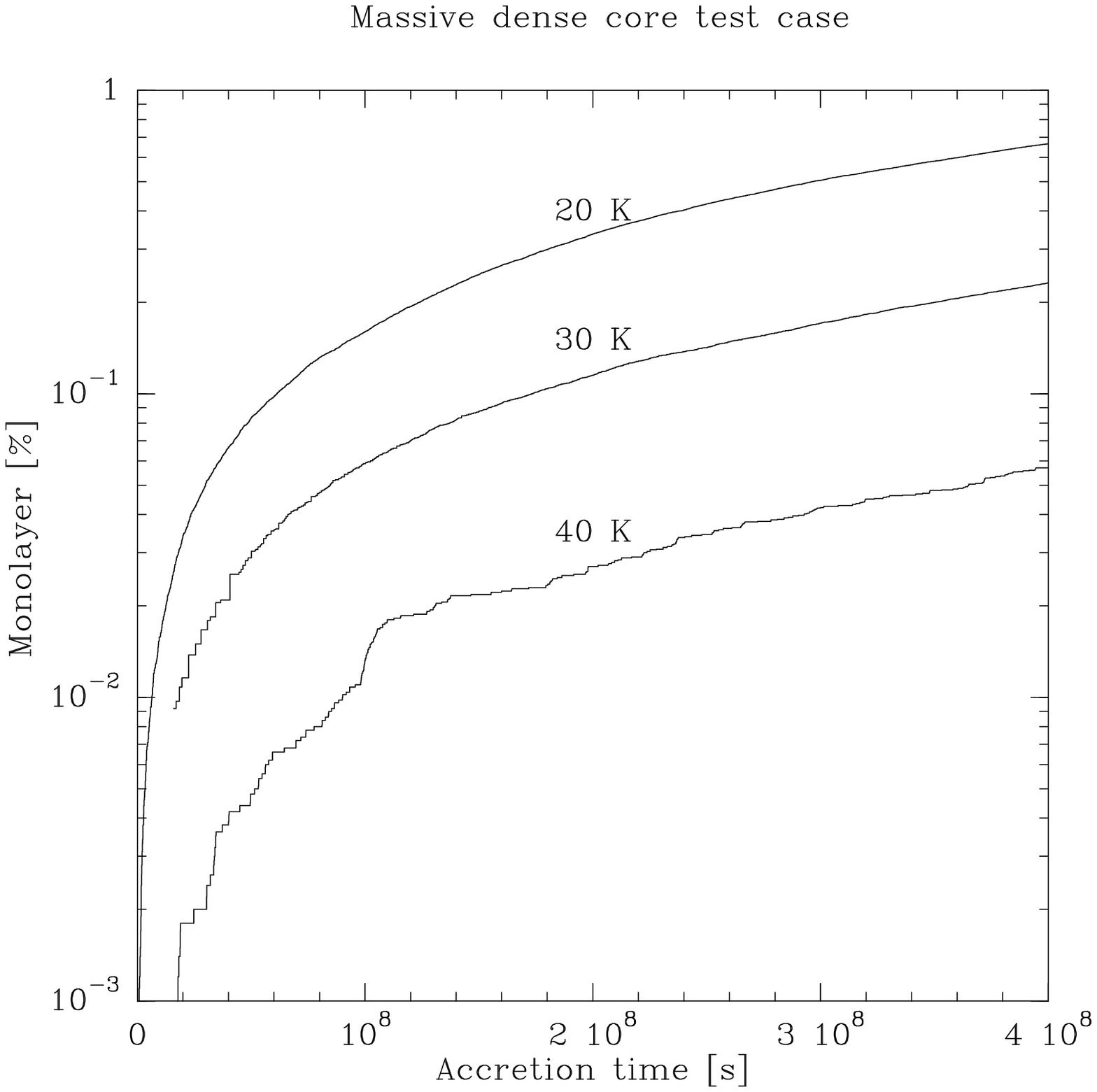}
 \includegraphics[width=220pt,angle=-90]{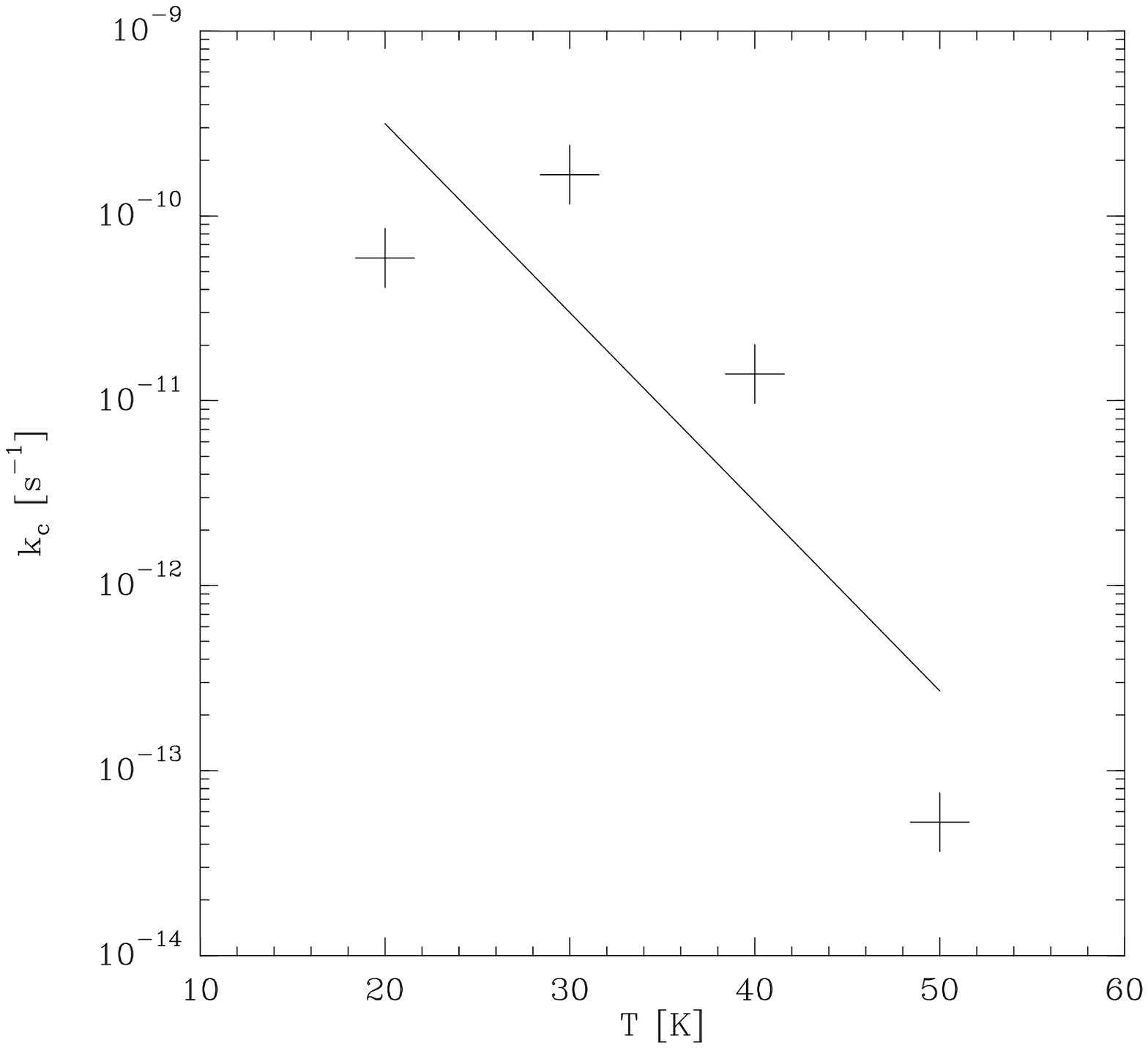}
\end{center}
 \caption{Plot of the bare-to-icy grain transformation rate $k_c$ as function of the temperature in the  high-mass protostellar objects considered.}
 \label{fig:kc-mass}
\end{figure}

\section{Impact of the main parameters}

The results of the previous section have been obtained for what we assume to be "typical parameters" in the considered proto-stellar environments. Although a clear difference between low-mass and massive protostars is present, the dependence of the snowborder caracteristics (extension and growth speed) on the main parameters should be explored. These main parameters, linked to the mixing effect which is mandatory for the snow border formation, are the stellar accretion rates and the turbulent velocity fields. 

Keeping the proto-planetary-disk and the massive protostar as basis for the following study, we now explore a larger range of value for $\varv_\mathrm{t}$ and $\dot{M}$. For proto-planetary-disks, the turbulent velocity field ranges between $0.01$ and $0.1$~\kms\ and the stellar accretion rate between $10^{-8}$ and $10^{-5}$~\msoly. Concerning high-mass protostellar objects, these values range between $1.0$ and $10$~\kms\ for the turbulent velocity field and between $10^{-5}$ and $10^{-2}$~\msoly\ for the accretion rate. 

\subsection{Impact on snow border growth}

The Fig.~\ref{fig:sbg} (color scale) describes the influence of the stellar accretion rate and the turbulence velocity on the growth of the snow border in AU\,yr$^{-1}$. The snow border growth ranges between 0 and 0.005~AU\,yr$^{-1}$ for the proto-planetary model and between 0 and 0.5~AU\,yr$^{-1}$ for the high-mass protostellar object model. In this sense, a null value means that the snow border formation will not occur. 
The snow border growth increases as stellar accretion rate decreases and turbulent velocity increases. On one hand, the stellar accretion prevents the diffusion of water molecules from the inner part of the objects to higher radii. On the other hand, turbulences support the diffusion towards the external parts. As a consequence, accretion rates lower than $10^{-5}$~\msoly\ (resp. $10^{-2}$~\msoly) allow the appearance of the snow border in protoplanetary disks (resp. HMPO). At the same time, turbulences higher than 0.2~\kms\ (resp. $2$~\kms) are required for these environments. 

\subsection{Impact on steady-state time}

Fig.~\ref{fig:sbg} (labelled white contour map) shows the time to reach the steady state equilibrium. This time increases strongly with accretion, and slightly with turbulent velocities: accretion slows down diffusion while turbulence improves it. The appearance of an extended snow border can only occur if its associated growth is important enough and the time to reach the steady-state is sufficient.

In the case of proto-planetary disks, the maximal extension for the snow border is obtained after 100~years for turbulence of 0.1~\kms\ and an stellar accretion rate of 5\ttp{-6}~\msoly, and reaches 0.2~AU (0.2~\% of the total size). This time is reasonnable compared to standard evolution time scale of a protoplanetary disk, meaning that the snow border exists. 

In the case of high-mass protostellar objects, the maximal extension can be obtained after 10$^5$~years, for a turbulent velocity field of 10~\kms\ and an accretion rate of 5\ttp{-3}~\msoly, and reaches 10\,000~AU (40~\% of the total size). Above this time, HMPOs are known to enter a phase of medium ionization due to the central star starting its life on the main sequence.

Nevertheless, the cases mentionned above are extreme compared to standard objects observed. More reasonnable value for the total size of the snow border should be around 0.04~AU for protoplanetary disks and 5000~AU for HMPOs. This values correspond to the lower left part of the graph in Fig.~\ref{fig:sbg}.

\begin{figure*}[t]
\begin{center}
 \includegraphics[width=220pt,angle=-90]{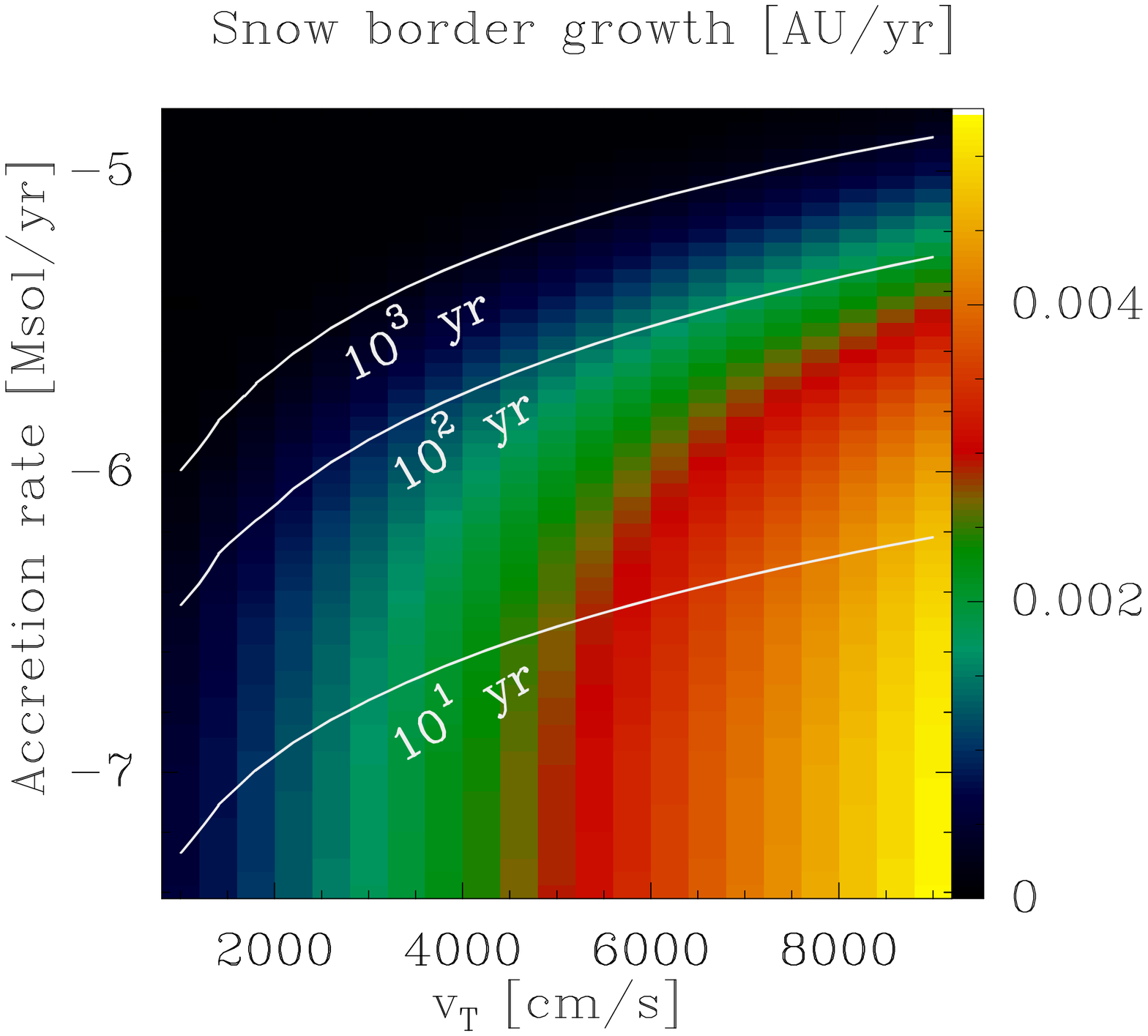}
 \includegraphics[width=220pt,angle=-90]{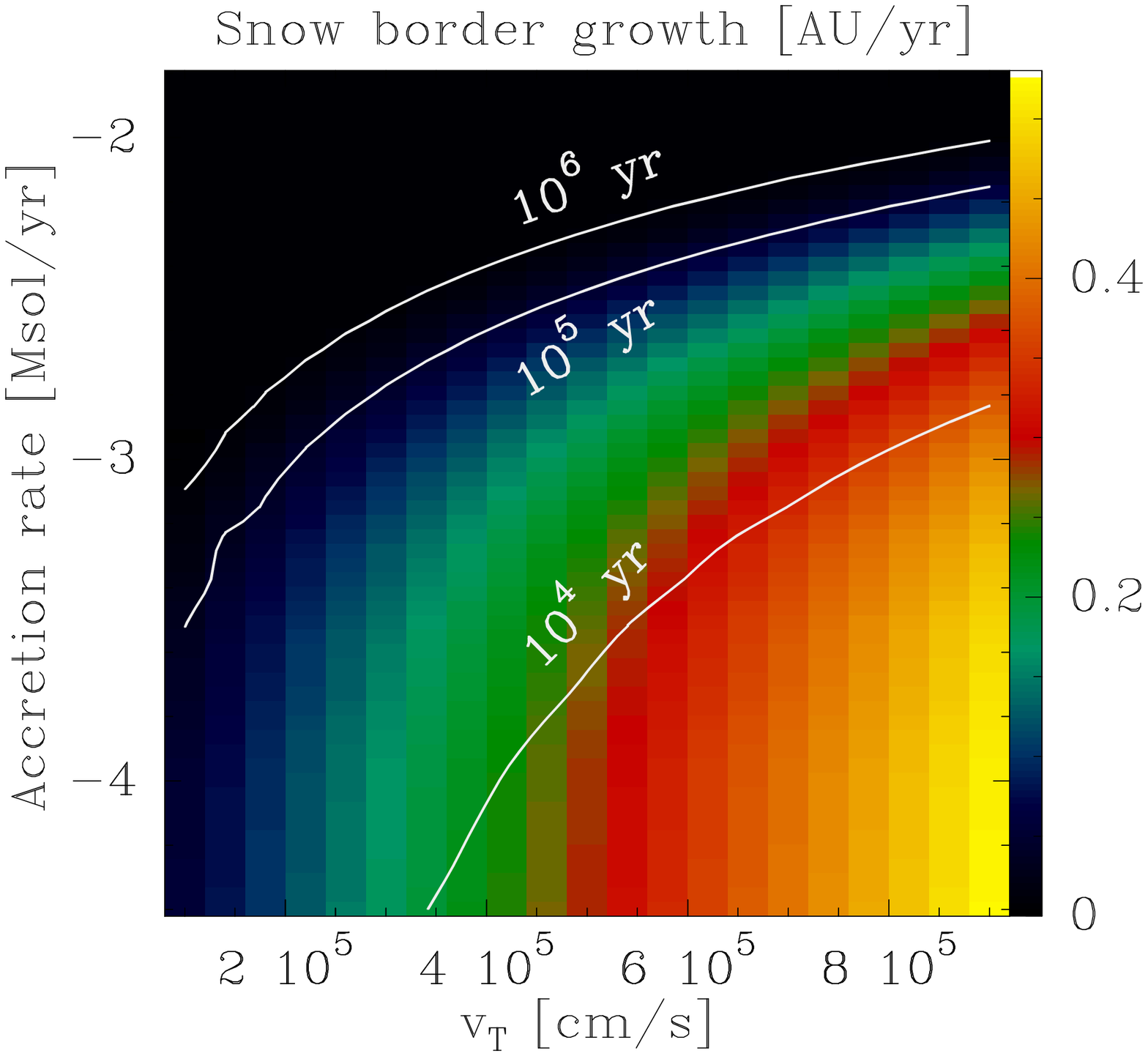}
\end{center}
 \caption{Overlay of the snow border growth (color scale) and the time to reach the steady-state (white contours) as functions of stellar accretion rates and turbulent velocity field in protoplanetary discs (left) and high-mass protostellar objects (right).}
 \label{fig:sbg}
\end{figure*}

\section{Discussion and conclusions}

The condensation of water ice under $100$~K requires long time scales. This allows the coexistence of both icy and bare grains in a zone beyond the snow line, that can be called the \textit{snow border}. This region can be significant compared to the size of the astrophysical objects considered in this work, particularly for massive dense cores where its extent can reach a fourth of the total system size.
More precisely, the existence of the snow border is triggered by the stellar accretion rate and the turbulent velocity field in the protostellar object. Our study shows that a minimal turbulent level of $\sim 0.1$~\kms\ in protoplanetary disks (resp. 1~\kms\ in HMPOs) is required to allow fast diffusion of water molecules. It also indicates that stellar accretion should be lower than $\sim 10^{-5}$~\msoly\ in protoplanetary disks ($\sim 10^{-2}$~\msoly\ in HMPOs) to allow an efficient appearance of the \textit{snow border} region, otherwise the water diffusion towards the object is prevented.  

The \textit{snow border} seems critical for observations of massive dense cores, where its size (few thousand AU at few kpc, hence $\sim 1$\arcsec) is not negligible. Indeed, to first order, the existence of such a region extends the zone in which molecules are present in the gas phase.
Interferometers, and obviously the ALMA project in the coming years, should be able to reach the spatial resolution corresponding to the \textit{snow border} extent. 
More precisely, the \textit{snow border} allows a non-negligible abundance of water in the gas phase, even at cold temperatures (down to $\sim 50$~K). 
This could be a critical point for the analysis of the upcoming water observations in massive dense cores by the Herschel Spatial Observatory.
However, we must stress that our study is based on carbon surface behaviours, whereas astronomical silicates react differently. 
Particularly, a fast condensation of water on silicate remains below $160$~K. As a consequence, a sharp release of complex molecules over 160~K is still valid, whereas the \textit{snow border} region will be characterised by a transfer of ice materials from carbon to silicate grains. In the future, this issue should be investigated, this discrepancy between the two type of grains being possibly important in the physical and chemical evolution of protostellar environments.

From a modelling point of view, this result opens new doors on several topics, and may have a non-negligible impact on the evolution of star forming regions. 
First, the \textit{snow border} existence allows a smooth transition of molecular abundances in the gas phase and on the grain surfaces. 
This is new compared to the abundance jump assumed commonly in chemical modellings or observation analyses. 
A natural next step will be to couple gas and grain chemistry to better constrain these modifications. 
In addition, as the formation of the first layer of ice depends on the grain size, it is of prime importance to study this aspect in further detail.
Concerning the modelling of strongly heated objects like of hot cores or hot corinos, the \textit{snow border} may modify the grain temperatures, hence the results obtained.
The coagulation of bare and large icy dust grains could also lead to the creation of new type of proto-planetary objects, through collisions between these two types of grains. 
Indeed, current modelling consider bare-bare or icy-icy grain interactions only \citep{ormel2007}. 
In addition, previous results obtained by \cite{kretke2007} or \cite{brauer2008} tend to show that formation of planetesimals is particularly enhanced in the neighbourhood of the snow line. 
Concerning the formation of planets, the fact that bare grains can exist at large distance from the central star could also play a role in the rocky vs gaseous planet formation issue, pushing its limit further away.\rm

\begin{acknowledgements}
The authors would like to thank the anonymous referee for a very constructive report, which greatly improved this manuscript. S. C. is supported by the Netherlands Organization for Scientific Research (NWO)
\end{acknowledgements}

\end{document}